\documentclass[aps,prl,reprint,superscriptaddress]{revtex4-1}
\usepackage{graphicx}
\usepackage{amsmath}
\usepackage{tabularx}

\begin{document}

\title{Coherent tunneling in an AlGaN/AlN/GaN heterojunction captured through an analogy with a MOS contact}

\author{Yannick Baines}
\email[e-mail: ]{yannick.baines@cea.fr}
\affiliation{Univ. Grenoble Alpes, F-38000 Grenoble, France}
\affiliation{CEA, LETI, MINATEC Campus, F-38054 Grenoble, France}

\author{Julien Buckley}
\affiliation{Univ. Grenoble Alpes, F-38000 Grenoble, France}
\affiliation{CEA, LETI, MINATEC Campus, F-38054 Grenoble, France}

\author{J\'er\^ome Biscarrat}
\affiliation{Univ. Grenoble Alpes, F-38000 Grenoble, France}
\affiliation{CEA, LETI, MINATEC Campus, F-38054 Grenoble, France}

\author{Gennie Garnier}
\affiliation{Univ. Grenoble Alpes, F-38000 Grenoble, France}
\affiliation{CEA, LETI, MINATEC Campus, F-38054 Grenoble, France}

\author{Matthew Charles}
\affiliation{Univ. Grenoble Alpes, F-38000 Grenoble, France}
\affiliation{CEA, LETI, MINATEC Campus, F-38054 Grenoble, France}

\author{William Vandendaele}
\affiliation{Univ. Grenoble Alpes, F-38000 Grenoble, France}
\affiliation{CEA, LETI, MINATEC Campus, F-38054 Grenoble, France}

\author{Charlotte Gillot}
\affiliation{Univ. Grenoble Alpes, F-38000 Grenoble, France}
\affiliation{CEA, LETI, MINATEC Campus, F-38054 Grenoble, France}

\author{Marc Plissonnier}
\email[e-mail: ]{marc.plissonnier@cea.fr}
\affiliation{Univ. Grenoble Alpes, F-38000 Grenoble, France}
\affiliation{CEA, LETI, MINATEC Campus, F-38054 Grenoble, France}

\date{\today}

\begin{abstract}
Due to their wide band gaps, III-N materials can exhibit behaviors ranging from the semiconductor class to the dielectric class.
Through an analogy between a Metal/AlGaN/AlN/GaN diode and a MOS contact, we make use of this dual nature and show a direct path to capture the energy band diagram of the nitride system. We then apply transparency calculations to describe the forward conduction regime of a III-N heterojunction diode and demonstrate it realizes a tunnel diode, in contrast to its regular Schottky Barrier Diode designation. Thermionic emission is ruled out and instead, a coherent electron tunneling scenario allows to account for transport at room temperature and higher.
\end{abstract}


\maketitle

\section{INTRODUCTION}
III-N materials are today at the heart of continuous academic and industrial efforts worldwide. With applications in lighting, radio-frequency or power technologies, Gallium, Indium and Aluminum Nitride alloys offer versatile and outstanding platforms that enable high performance electronics and enhanced solutions in multiple sectors.

Due to its intrinsic properties, GaN is nowadays drawing a lot of attention in the power electronics area as it leads to lower conversion losses at higher frequency compared to Silicon or Silicon Carbide devices. Multiple leading companies have already demonstrated GaN power technologies with extremely promising results \cite{Kaneko15, Moens14, Kim13, Jones16, LB},  thereby paving the way to new and greener energy converters with a market expected to ramp up in the forthcoming years. 

At the core of most of these GaN power devices lies the AlGaN/GaN heterojunction due to its exceptional properties, in particular the existence of a two-dimensional electron gas. Finding its origin in spontaneous and piezoelectric polarization \cite{Ambacher99}, it gives rise to a high electron concentration combined with a high electron mobility. 

In this work, we propose to draw parallels between an AlGaN/AlN/GaN heterojunction diode and a Metal/Oxide/Semiconductor (MOS) contact by using the duality found in III-N wide band gap materials: semiconductors on the one hand \cite{Huang06, Yoshizumi07, Kizilyalli14}, dielectrics on the other hand \cite{Moens15, Moereke16, Selvaraj07}, which can depend on their doping and on their electromagnetic environment. With the use of this analogy, we are able to map the energy band diagram at the rectifying contact vicinity in a direct way in order to address the underlying transport mechanisms. By combining the Transfer Matrix formalism \cite{Ando87} to compute the system's transparency and the Tsu-Esaki current formula \cite{Duke69, Tsu73}, we describe the forward conduction regime of the heterojunction diode with respect to applied voltage and operating temperature.

We show thereby that the III-N heterojunction diode realizes a tunnel diode and rule out thermionic emission, too often incorrectly used to capture such architectures. Through this approach we propose an alternative path of understanding the physics of III-N heterostructures and devices.

\section{MOS analogy}
We start by considering an AlGaN/AlN/GaN double heterojunction, a widely used design \cite{Wang07, Shen01, Ohmaki06} where the AlGaN layer is often referred to as the barrier layer, the AlN as the spacer layer and the GaN as the channel layer which contains a Two-Dimensional Electron Gas (2DEG) at the AlN/GaN interface. The system is completed with a gate metal on top of the AlGaN, more specifically Titanium Nitride (TiN). A common approach to studying this system is the use of a Poisson-Schr\"{o}dinger simulator which allows the derivation of the energy band diagram of the contact and captures the existence of the 2DEG \cite{Zhang14}. We propose here an alternative approach to the problem.

\begin{figure}
\includegraphics[width=8.5cm]{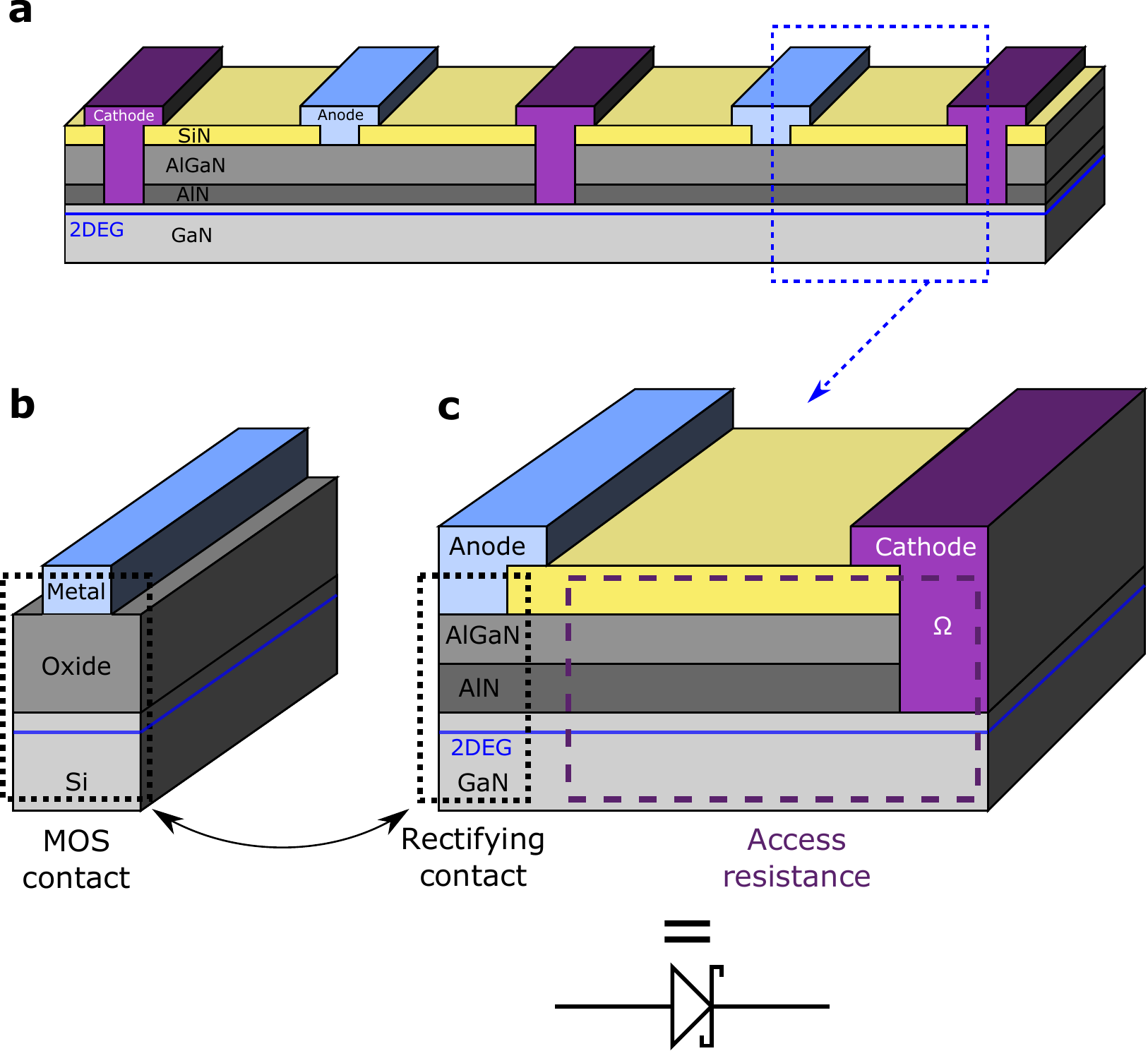}%
\caption{\textbf{a}: AlGaN/AlN/GaN interdigitated heterojunction diode. \textbf{b}: Typical MOS contact as found in a Silicon MOS capacitor. \textbf{c}: AlGaN/AlN/GaN diode half cell separated into two main components: the rectifying contact and the access resistance.}
\label{Fig1}
\end{figure}

To allow an equivalence between the TiN/AlGaN/AlN/GaN contact and a MOS contact we make two main assumptions. First we assume that the AlGaN and AlN layers are depleted from free carriers, thus allowing to consider purely their dielectric nature and equate them to the gate oxide found within a MOS. Then we assume the GaN channel layer to be a n type semiconductor which will be the stage of an electron accumulation layer as found at the semiconductor/oxide interface of a MOS contact. Figure \ref{Fig1} illustrates this analogy by representing schematically an AlGaN/AlN/GaN diode (\textbf{a}) where the rectifying contact part or anode (\textbf{b}, dotted box) can be mapped on an effective MOS contact (\textbf{c}).

This analogy allows us to employ the main equation used to describe MOS capacitors with a first focus being the derivation of the energy band diagram. We start by considering only the rectifying contact part and express the potential equilibrium as found in a MOS capacitor:
\begin{equation}
V=V_{fb}+V_{ox}+\Psi_s
\label{eq1}
\end{equation}
Where $V$ is the metal potential (referenced to the bulk GaN supposed grounded), $V_{fb}$ the flat band potential, $V_{ox}$ the potential drop across the oxide and $\Psi_s$ the surface potential of the semiconductor which translates the amount of band bending in the accumulation layer.

We detail now the different terms of this equation when applied to the Metal/AlGaN/AlN/GaN contact.
The flat band potential expresses the difference between the Fermi levels in the metal (TiN) and in the semiconductor (bulk GaN).
\begin{equation}
V_{fb}=W-\chi ^{GaN}-(E_C-E_F)^{GaN}
\label{eq2}
\end{equation}
Where $W$ is the gate metal work function, $\chi$ the electron affinity of GaN and $(E_C-E_F)$ the energy difference between the bottom of the conduction band and the Fermi energy in the bulk GaN. 

\begin{figure}
\includegraphics[width=7.5cm]{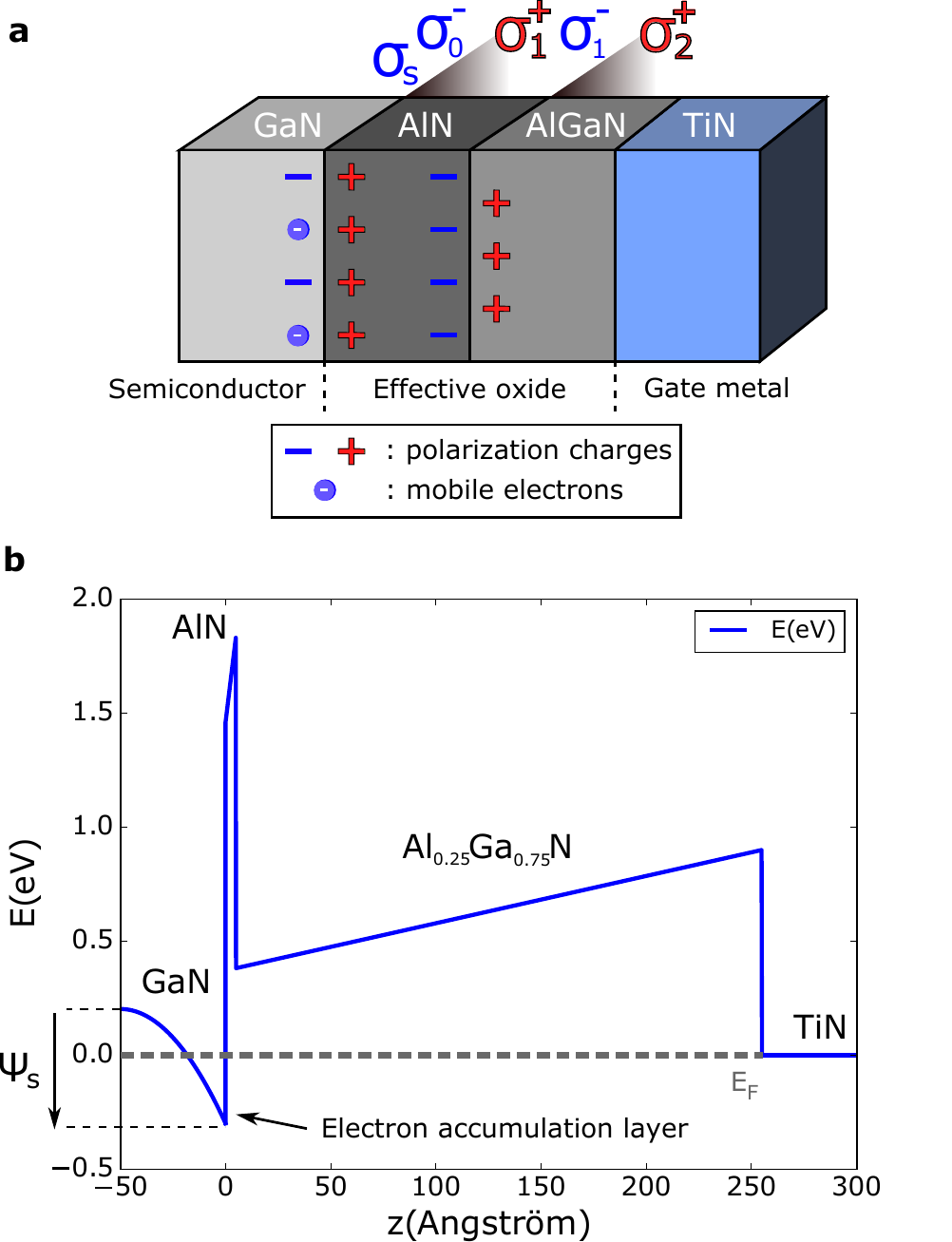}%
\caption{\label{Fig2}\textbf{a}: GaN/AlN/AlGaN/TiN stack showing the different interface charges (fixed and mobile) found at each heterojunction. \textbf{b}: Reference structure computed conduction band profile employing the MOS analogy.}
\end{figure}

As far as $V_{ox}$ is concerned, we recall that III-N materials have polar bonds. Within the bulk, the dipoles created by each chemical bond are canceled by the neighboring ones, two by two. However, at a hetero-interface this balance is no longer present and polarization surface charges develop at them. Figure \ref{Fig2} (\textbf{a}) illustrates the different interface charges and their polarity present in the case of a Ga-face oriented heterojunction. The effective oxide (AlGaN+AlN) under consideration is therefore natively charged.

By taking into account spontaneous and piezoelectric charges and expressing the continuity of the normal component of the electric field in the structure, the voltage drop across the AlN and AlGaN layers reads:
\begin{equation}
V_{ox}=-\frac{\sigma_s+\sigma_{10}}{C_\mathrm{{AlN}}}-\frac{\sigma_s+\sigma_{10}+\sigma_{21}}{C_{\mathrm{AlGaN}}}\\
\label{eq3}
\end{equation}
Where $\sigma_s$ denotes the accumulation charge (mobile electrons) per unit area in the GaN accumulation layer and $C_{\mathrm{AlN}}$ and $C_{\mathrm{AlGaN}}$ denote the equivalent AlN and AlGaN capacitances per unit area respectively. $\sigma_{10}=\sigma_0^-+\sigma_1^+$ and $\sigma_{21}=\sigma_1^-+\sigma_2^+$ express the net polarization charges found at the GaN/AlN and AlN/AlGaN interfaces respectively (refer to supplementary information). For each material, the spontaneous contribution is determined following \cite{Ambacher99} and the piezoelectric part is determined following \cite{Ioffe} by assuming the AlN and the AlGaN to be fully strained on a relaxed GaN layer.

It can be shown that in the case of a 2D accumulation layer, the surface charge density $\sigma_s$ can be related to the surface potential $\Psi_s$ by solving the Poisson equation \cite{Seiwatz58}:
\begin{equation}
\sigma_s=\pm\frac{\epsilon k_BT}{qL_D}\sqrt{\frac{N_v}{p_0}F^-+\frac{q\Psi_s}{k_BT}+\frac{n_0}{p_0}\left(\frac{N_c}{n_0}F^+-\frac{q\Psi_s}{k_BT}\right)}
\label{eq4}
\end{equation}
Where $\sigma_s$ is taken to be negative when $\Psi_s$ is positive and vice versa.
$n_0,p_0$ correspond to the electron and hole concentrations in the bulk GaN, $N_C$ and $N_V$ to the GaN conduction and valence band effective density of states and $L_D$ to the Debye length written as:
\begin{equation}
L_D=\sqrt{\frac{\epsilon k_BT}{2q^2p_0}}
\label{eq5}
\end{equation}
The $F^-$ and $F^+$ functions are defined as:
\begin{align}
F^-=F_{3/2}\left(\frac{E_V-E_F-q\Psi_s}{k_BT}\right)-F_{3/2}\left(\frac{E_V-E_F}{k_BT}\right) \\
F^+=F_{3/2}\left(\frac{E_F-E_C+q\Psi_s}{k_BT}\right)-F_{3/2}\left(\frac{E_F-E_C}{k_BT}\right)
\label{eq6}
\end{align}
With $F_{3/2}$ corresponding to the normalized Fermi-Dirac integral $F_j(x)$ with order $j=3/2$ \cite{Aymerich83}:
\begin{equation}
F_j(x)=\frac{1}{\Gamma (j+1)}\int_{0}^{\infty}\frac{\xi^j}{1+\mathrm{exp}(\xi-x)}d\xi
\label{eq8}
\end{equation}
Combining (\ref{eq3}) and (\ref{eq4}) and injecting the outcome with (\ref{eq2}) into (\ref{eq1}), one can solve numerically the surface potential $\Psi_s$ at the GaN/AlN interface, which in turn determines $\sigma_s$ and finally $V_{ox}$.

To illustrate the procedure, we apply it to the computation of the band diagram of a reference structure and more precisely we focus on the conduction band profile. The main operating and structural parameters used are summarized in the following table:
\begin{table}[ht]
\centering
\caption{Reference structure main parameters}
\begin{tabular*}{0.45\textwidth}{@{\extracolsep{\fill}}lr}
\hline\hline\\
Metal voltage ($V$) & 0 V \\
Temperature & 300 K \\
TiN work function ($W$) & 4.7 eV \cite{Li14} \\
$\mathrm{Al}_{0.25}\mathrm{Ga}_{0.75}$N thickness & 25 nm \\
AlN thickness & 0.5 nm \\
Bulk GaN electron concentration ($n_0$) & 1e15 cm$^{-3}$ \\\\
\hline\hline
\end{tabular*}
\label{tab1}
\end{table}

Figure \ref{Fig2} (\textbf{b}) illustrates the obtained conduction band profile for the reference TiN/AlGaN/AlN/GaN contact where the conduction band offsets were accounted for following \cite{Ambacher99}. Note that the profile of the conduction band from the bulk GaN (at $z=-50$ A) to the GaN/AlN interface ($z=0$ A) is shown for the sake of clarity and was not calculated explicitly. Indeed the model used assumes a 2D accumulation layer and returns the amount of band bending at the surface of the GaN referenced to the bulk, where the position of the Fermi energy is known through $n_0$.

We observe that the existence of the 2D electron gas, or the electron accumulation layer, naturally arises from the calculation and is induced by the strong polarization charges found within the hetero-structure. The extracted sheet carrier density within the accumulation layer is $n_s=1e13$ cm$^{-2}$ which agrees well with experimental values reported for similar heterojunctions \cite{Wang07, Shen01}.

\section{Practical considerations}
\subsection{Transfer length}
\begin{figure}
\includegraphics[width=8.5cm]{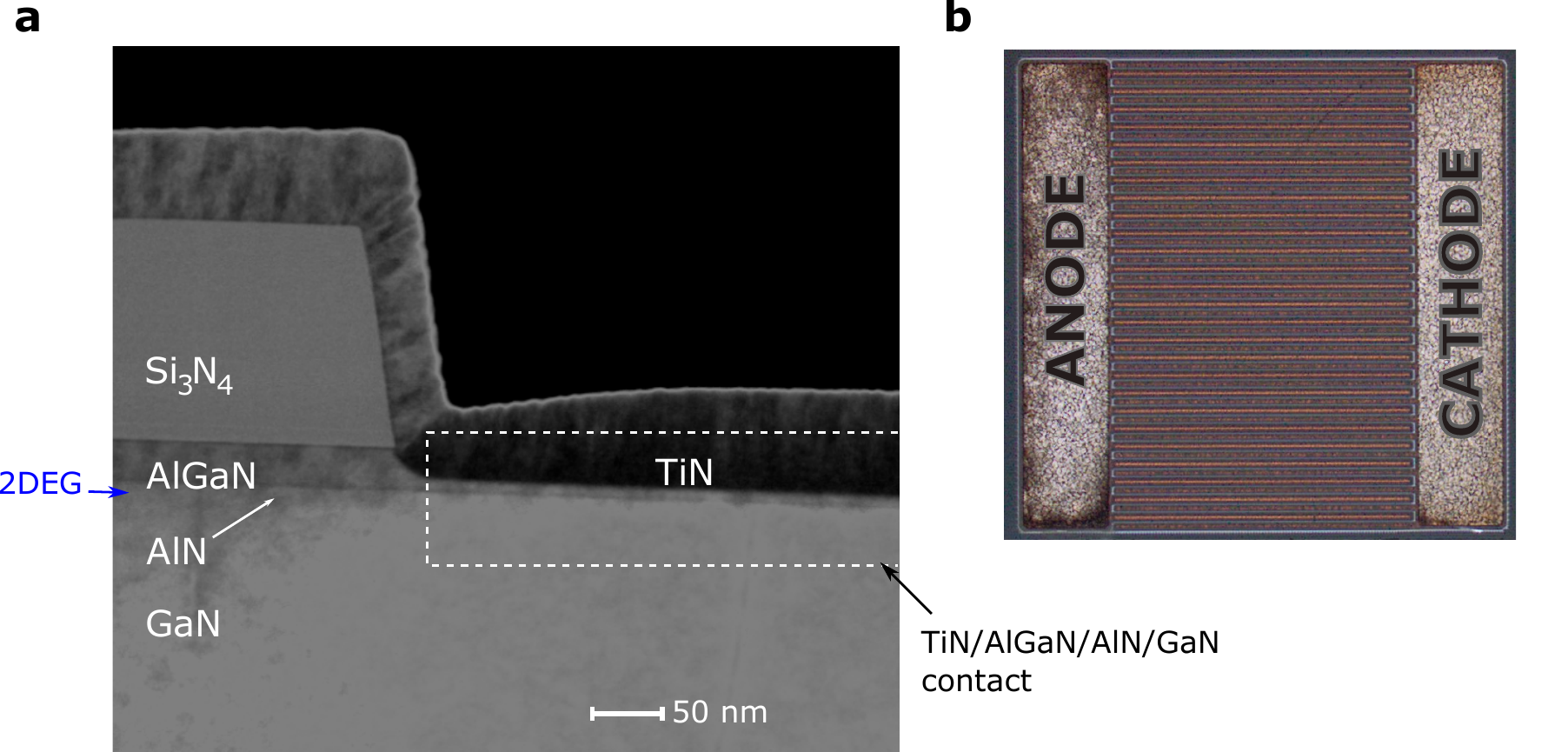}%
\caption{\label{Fig3} \textbf{a}: TEM close view around the anode contact found on a III-N heterojunction diode fabricated at CEA-LETI. \textit{Dashed box}: TiN/AlGaN/AlN/GaN  contact. \textbf{b}: Upper view of the power diode showing the interdigitated designed used.}
\end{figure}

III-N heterojunction diodes have been fabricated at CEA-LETI using 200 mm GaN on Silicon wafers grown on site that were processed through a CMOS compatible integration flow. Figure \ref{Fig3} shows a close view of the TiN/AlGaN/AlN/GaN contact found on a  processed diode which in this case contains a recessed $\mathrm{Al}_{0.25}\mathrm{Ga}_{0.75}$N barrier with 6 nm left above the AlN spacer. For the purpose of this study, the fabricated diodes were designed with anode length, cathode length and anode to cathode distance all of 15 $\mu$m.

Before going into the details of the transport mechanisms entering the on state of the presented diode, we recall a common feature found in contact physics: the transfer length notion. This parameter, often characterized through Transmission Line Measurements (TLM), expresses the length over which a contact, ohmic in the TLM case, is effectively operating. The concept can be extended to nonlinear contacts and corresponding nonlinear transmission lines \cite{Piot11}. Understand thereby that for heterojunction diodes with sufficiently long anode contacts, the majority of the current is emitted at the anode periphery in the forward regime.
\begin{figure}
\includegraphics[width=7.5cm]{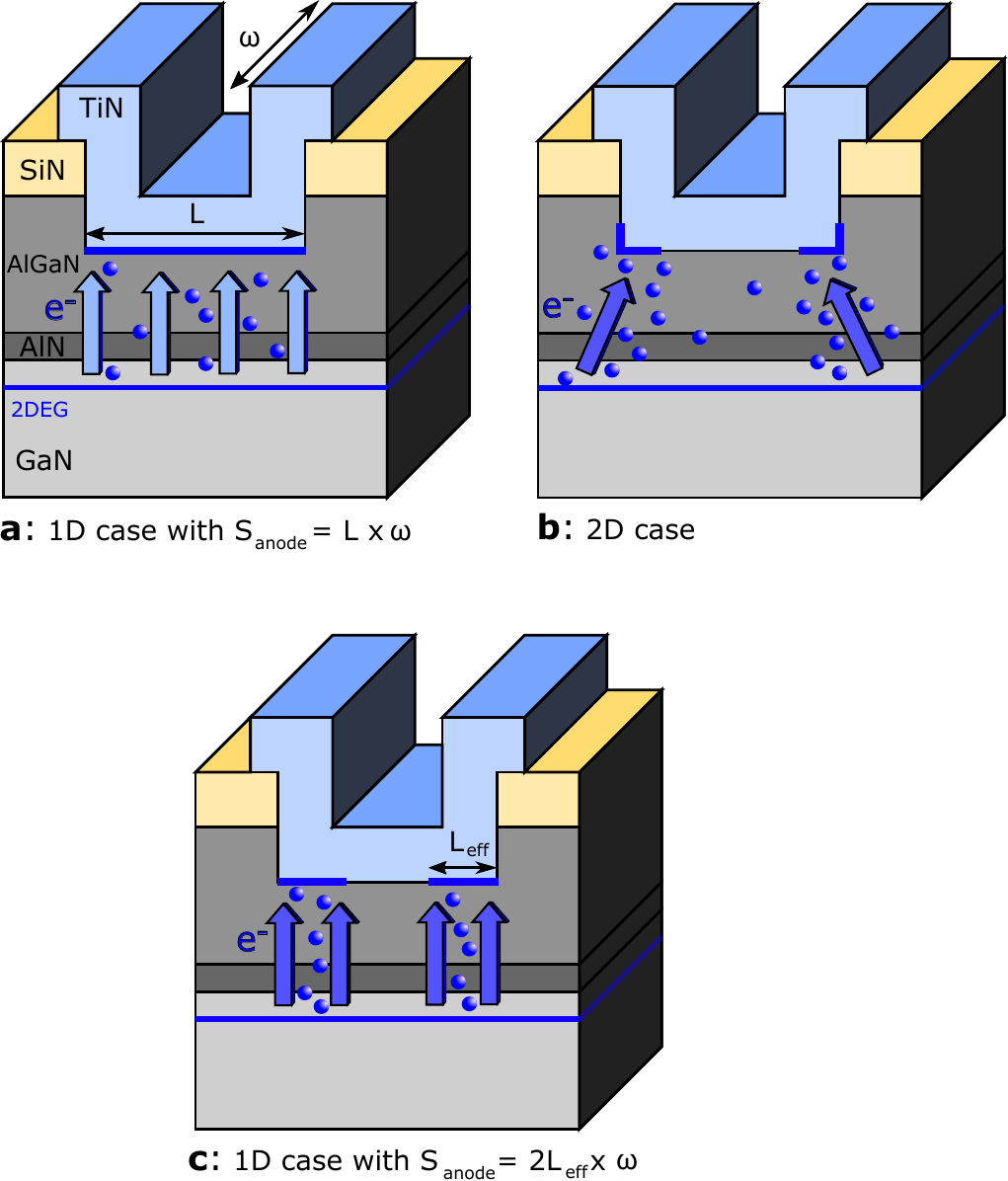}%
\caption{\label{Fig4}Electron flow under the anode contact. \textbf{a}: 1D standard case, the current is assumed constant over the entire anode length. \textbf{b}: 2D case, the current varies over the anode length with the majority of the current emitted at the periphery. \textbf{c}: 1D intermediate case, an effective length is introduced at the edges of the anode and over which the current is assumed constant.}
\end{figure}

To account for this feature and avoid 2D calculations to reproduce the on state current, we propose a compromise between the regular 1D surface independent current emission (figure \ref{Fig4}, \textbf{a}) and the more complex 2D surface dependent current (figure \ref{Fig4}, \textbf{b}). Figure \ref{Fig4} (\textbf{c}) represents the intermediate case we will consider in the following which makes use of an effective anode contact length $L_{eff}$ over which the current is assumed constant. Through this simplification a 1D approach is possible with a reduced contact surface $S=2L_{\mathrm{eff}}\omega$, where $\omega$ stands for the anode width.

\subsection{Anode recess impact}
As detailed in the previous section, the rectifying contact of the diode under study is realized through a partial recess of the $\mathrm{Al}_{0.25}\mathrm{Ga}_{0.75}$N barrier with a targeted thickness of 6 nm left above the AlN spacer. This strategy is common in such architectures and offers the benefit of reducing the turn-on voltage of the diode \cite{Bahat12, Zhu15, Hu14} allowing more current to be delivered in the on state, a critical feature for end users.

We derive the band diagram of the recessed diode using the same set of parameters as in the reference structure (table \ref{tab1}) and adjust the AlGaN thickness to 6 nm. The calculated resulting sheet carrier density under the contact is $n_s=7.3e12$ cm$^{-2}$. C(V) measurements performed on representative test structures reveal an experimentally smaller value falling in the range $1e12$ cm$^{-2}$ to $3e12$ cm$^{-2}$.

To explain this difference, we turn back to the etching process steps which allow the anode contact recess. The process sequence uses Reactive Ion Etching and comprises the Si$_3$N$_4$ passivation layer opening using Fluorine species (F$^-$) followed by the $\mathrm{Al}_{0.25}\mathrm{Ga}_{0.75}$N etching using Chlorine species (Cl$^-$). Fluorine implantation through a similar process was already reported in the AlGaN barrier by coworkers \cite{Lehmann15} and we believe that within the etching condition used, negatively charged ions contaminate the active layers, more precisely the AlGaN left under the TiN metal.

In order to take into consideration ion trapping in the AlGaN, we introduce an effective surface charge at the AlN/AlGaN interface. We modify $\sigma_{21}$ in equation (\ref{eq3}) to $\sigma_{21}=\sigma_1^-+\sigma_2^++\sigma_{\mathrm{etch}}$ where $\sigma_{\mathrm{etch}}$=-0.013 C.m$^{-2}$ represents a negative surface charge density linked to Fluorine/Chlorine ion implantation. The resulting sheet carrier density becomes $n_s$=1.8e12 cm$^{-2}$ which falls into the experimental range.
 
\section{Tunneling current}
Once the conduction band profile is known, the question of the current flowing through the contact can be addressed through the transmission probability of an electron moving from the accumulation layer to the metal. In order to determine this probability through an arbitrary potential barrier, in our case the conduction band profile of the AlN and AlGaN layers, we employ the Transfer Matrix formalism \cite{Ando87} which goes beyond the WKB approximation \cite{Wentzel26, Kramers26, Brillouin26}.

\begin{figure}
\includegraphics[width=8.5cm]{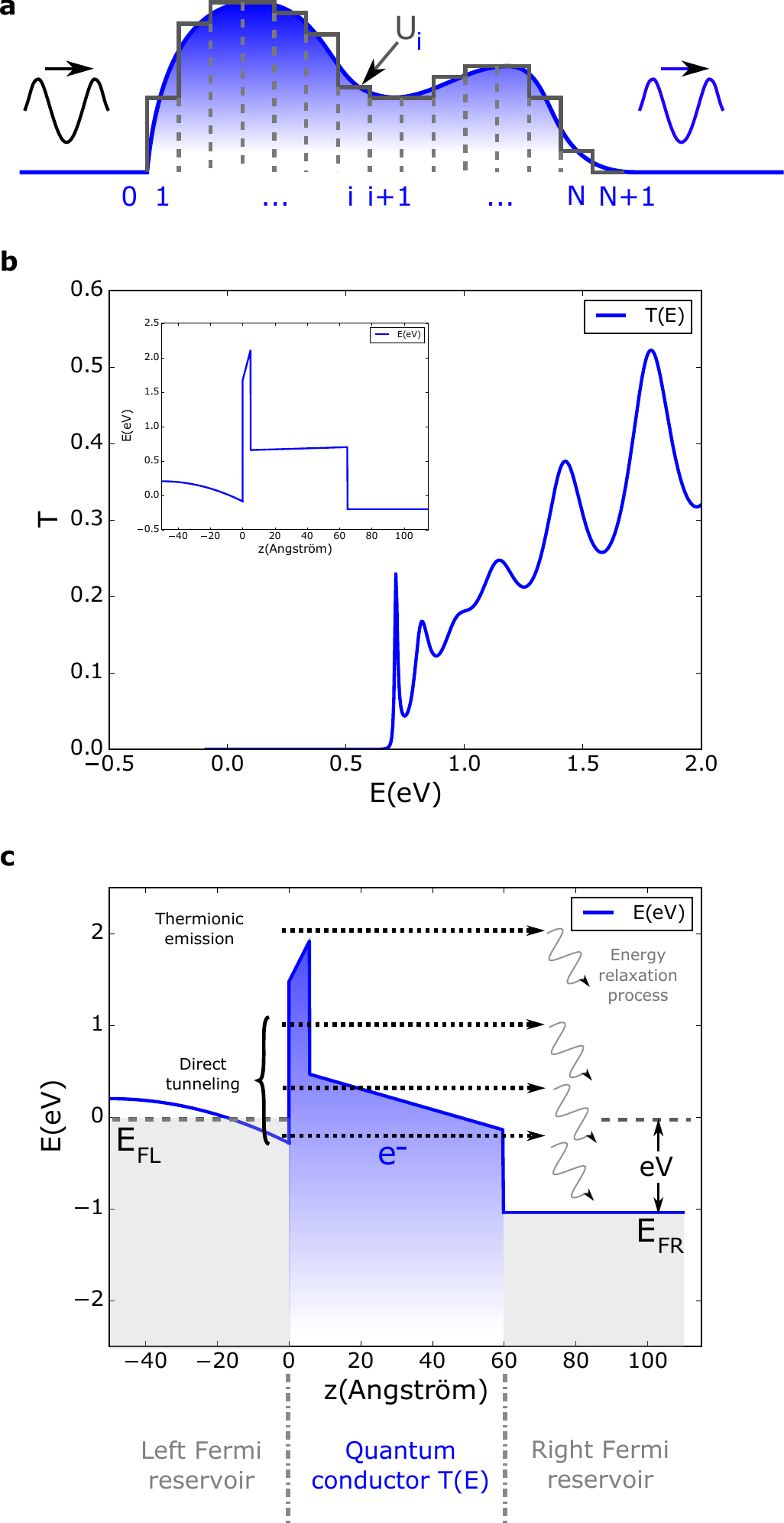}%
\caption{\label{Fig5}\textbf{a}: Arbitrary potential barrier discretization. \textbf{b}: Transmission probability as a function of energy at $V$=200 mV and T=300 K calculated using the transfer matrix formalism. \textit{Inset}: Associated recessed anode conduction band profile. \textbf{c}: Schematic representation of various electron tunneling events accounted for in the recessed anode diode using the Tsu-Esaki formula.}
\end{figure}

The procedure to follow is to discretize the potential barrier into N regions of constant potential (figure \ref{Fig5}, \textbf{a}) where locally the shape of the electron wavefunction $\Phi_i$ in region $i$ can be assigned as:
\begin{equation}
\Phi_i(z)=A_i\mathrm{exp}(jk_iz)+B_i\mathrm{exp}(-jk_iz)
\label{eq9}
\end{equation}
Where the wavevector as a function of energy $E$, potential $U_i$ and effective mass $m^*$ reads:
\begin{equation}
k_i=\sqrt{[2m_i^*(E-U_i)]}/\hbar
\label{eq10}
\end{equation}
Through the continuity of the wavefunction from the $i$ to the $i+1$ region, one can express how it transforms through a local transfer matrix M$_i$:
\begin{align}
 \label{eq11}
    M_i &=
    \begin{bmatrix}
    (1+S_i)\mathrm{e}^{-j(k_{i+1}-k_i)z_i} & (1-S_i)\mathrm{e}^{-j(k_{i+1}+k_i)z_i} \\
    (1-S_i)\mathrm{e}^{j(k_{i+1}+k_i)z_i} & (1+S_i)\mathrm{e}^{j(k_{i+1}-k_i)z_i} \\
    \end{bmatrix} \\
 \label{eq12}
    S_i &=\frac{m_{i+1}^*k_i}{m_i^*k_{i+1}}	
\end{align}
By iteration, a product of N matrices leads to the global transfer matrix through the potential barrier. From it, the electron transmission probability T(E) from region $i=0$ (accumulation layer) to region $i=N+1$ (metal) can be derived as a function of energy:
\begin{align}
 \label{eq13}
    M_i &=\prod_{i=0}^{i=N}M_i=
    \begin{bmatrix}
    M_{11} & M_{12} \\
    M_{21} & M_{22} \\
    \end{bmatrix} \\
 \label{eq14}
    A_{N+1}&=\frac{m_{N+1}^*k_{0}}{m_0^*k_{N+1}}\frac{1}{M_{22}}\\
 \label{eq15}
 	T(E)&=\frac{m_0^*k_{N+1}}{m_{N+1}^*k_0}|A_{N+1}|^2
\end{align}

We show in figure \ref{Fig5} (\textbf{b}) the calculation of the transmission probability T(E) for the recessed TiN/AlGaN/AlN/GaN contact. The computation is carried out from the bottom of the conduction band in the accumulation layer to a cut-off energy of 2 eV, using a voltage bias of 200 mV at 300 K.

In order to derive the current flow through the contact, we follow the Landauer vision of transport in a mesoscopic system where its quantum resistance is expressed in terms of the transmission and reflection properties of the structure \cite{Landauer70, Lenstra88,Meir92, Jauho94}. In the coherent tunneling limit and none-interacting electrons, the current density reads:

\begin{equation}
J(V)=\frac{2e}{h}\int T(E,V)\big[f_L(E)-f_R(E)\big]dE
\label{eqLandauer}
\end{equation}
Where $f_L$ and $f_R$ represent the equilibrium distribution functions in both reservoirs, indexed $L$ and $R$, connected to the quantum conductor and $T(E,V)$ denotes the conductor's transmission probability.

The use of such an approach to mesoscopic transport has proved to be versatile and a powerful tool which has been applied to various systems such as molecular junctions and transistors \cite{Solomon10,Guedon12, Ying16}, carbon nanotubes \cite{Wu07}, or quantum dots \cite{Baines12}.

In the case of parallel wavevector conservation and effective masses for each parabolic band of interest, equation (\ref{eqLandauer}) can be reduced to the so called Tsu-Esaki current formula \cite{Duke69, Tsu73}:

\begin{equation}
J(V)=\frac{qm_0^*k_BT}{2\pi^2\hbar^3}\int_{E_{0}}^{\infty}T(E_{\perp},V)\mathrm{ln}\Bigg[\frac{1+e^{\frac{E_F-E_{\perp}}{k_BT}}}{1+e^{\frac{E_F-qV-E_{\perp}}{k_BT}}}\Bigg]dE_{\perp}
\label{eq16}
\end{equation}

Where $T(E_\perp,V)$ is the transmission probability at transverse energy, that is to say in the transport direction, and which will be evaluated through the Transfer Matrix formalism. The logarithmic term arises from the assumption that the occupancy functions $f_{L(R)}$ follow the Fermi-Dirac distribution in both leads. The metal (TiN) and the electron accumulation layer (2DEG) are treated here as incoherent reservoirs where temperature dictates the precise shape of these distributions. The summation from the bottom of the conduction band ($E_0$) at the GaN/AlN interface to a high energy cut-off allows to account for electron emission from the GaN accumulation layer at different energies. Therefore multiple elastic processes, from direct tunneling in the forbidden band gap of the effective oxide (AlN and AlGaN) to thermionic emission over the potential barrier (above the AlN), are taken into consideration which is represented schematically in figure \ref{Fig5} (\textbf{c}). Within this framework, an electron emitted at energy E from the 2DEG (left Fermi reservoir) tunnels coherently to the contact Metal (right Fermi reservoir) where it is absorbed. Understand thereby that it losses its energy and previous state memory through incoherent relaxation.

\section{Application to forward conduction}
We focused in the last sections on the rectifying contact part of the lateral diode (figure \ref{Fig1} \textbf{b}, dotted box). In order to reproduce the I(V) characteristics of the entire device, we also need to take into consideration the access resistance (figure \ref{Fig2}, \textbf{b} dashed box), that is to say the 2DEG resistance in the anode to cathode region in series with the ohmic contact's resistance. We suppose this resistance $R$ to be constant at a given temperature within the voltage bias range we are about to explore. The effect of the access resistance is accounted for by setting the effective voltage drop across the rectifying contact in the right hand side of equation (\ref{eq16}) to: $V_{\mathrm{eff}}=V-RI$, $V$ being now the Anode-Cathode voltage bias. The diode's current density then becomes implicit and needs to be solved at each applied bias. Finally the total current is determined by recalling that $I(V)=2\omega L_{\mathrm{eff}} J(V)$ (figure \ref{Fig4}, \textbf{c}), where $\omega$=26 mm by design.

\begin{figure}
\includegraphics[width=8.5cm]{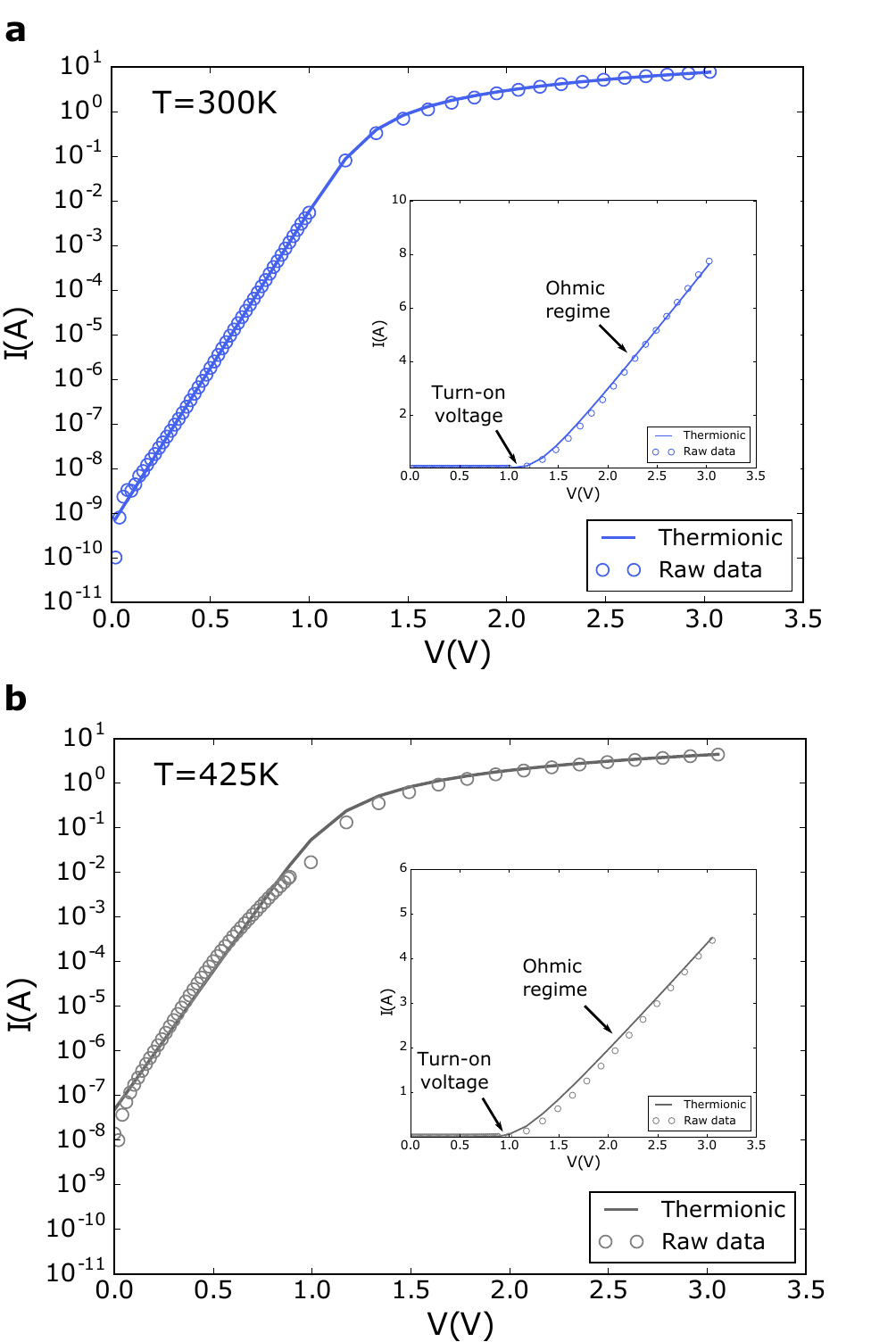}%
\caption{\label{Fig6} \textbf{a}: Semilog plot of the I(V) on state characteristics of the recessed diode at 300 K fitted using the thermionic formula. \textit{Inset}: Linear plot showing the ohmic regime above the turn-on voltage. \textbf{b} and \textit{inset}: 425 K case.}
\end{figure}

Before presenting the results of our calculations we recall a widely used empirical formula to reproduce the on state current of several different types of diodes including the studied architecture, the thermionic formula:
\begin{equation}
I_{TE}(V)=SA^*T^2\mathrm{exp}\left(\frac{-q\Phi_b}{k_BT})\right)\mathrm{exp}\left(\frac{q(V-RI)}{\eta k_BT}\right)
\label{eq17}
\end{equation}

\begin{figure*}
\includegraphics[width=17.5cm]{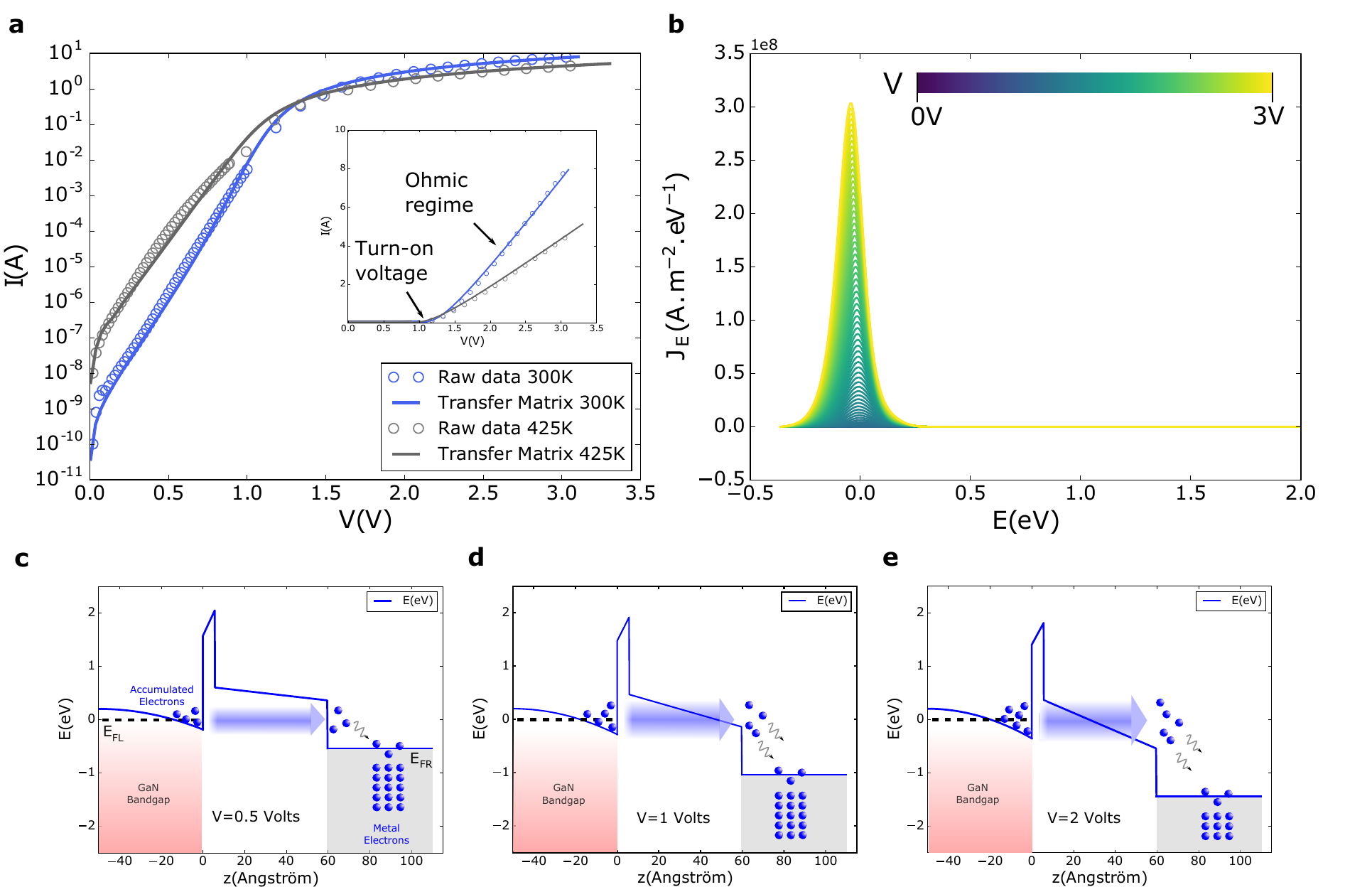}%
\caption{\label{Fig7} \textbf{a}: Semilog plot of the I(V) characteristics of the recessed diode at 300 K and 425 K reproduced using the model exposed. \textit{Inset}: Linear plot showing the ohmic regime above the turn-on voltage at both temperatures. \textbf{b}: Current density per unit area and energy as a function of energy at all voltages at 300 K. $E=0$ denotes the Fermi energy in the bulk GaN. \textbf{c to e}: Calculated conduction band profiles of the rectifying contact at room temperature and for applied biases of 0.5, 1 and 2 V. The shaded area in the accumulation layer represents the main energies contributing to the total current. Note that the contact sees an effective voltage drop of $V-RI$ which is noticeable above the turn-on voltage, that is to say case \textbf{e} in the present situation.} 
\end{figure*}

Where $S$ denotes the rectifying contact surface, $A^*$ denotes the modified Richardson constant, $\Phi_b$ the barrier height and $\eta$ the ideality factor. Figure \ref{Fig6} shows the forward conduction state of the fabricated diode at room temperature fitted via equation (\ref{eq17}) at 300 K (\textbf{a}) and 425 K (\textbf{b}). The adjustable parameters extracted from each fit are ($\Phi_b$=0.76 eV, $\eta$=2.39, $R$=0.21 $\Omega$) at room temperature and ($\Phi_b$=0.94 eV, $\eta$=1.91, $R$=0.4 $\Omega$) at 425 K. Although the agreement seems very good, such fits miss completely the physics of the device which we are about to show. On the one hand high ideality factors are synonymous of underlying tunnel events \cite{Sze06} which thermionic emission does not account for. On the other hand  $\Phi_b$ and $\eta$ need to have strong temperature dependencies in order to capture the evolution of the current, another sign of the irrelevance of such a physical picture in the present case.

Figure \ref{Fig7} (\textbf{a}) presents the results of the calculations compared to the raw data acquired at 300 K and 425 K. As can be observed, an excellent agreement is met within temperature and over the entire operating voltage range with the current spanning over more than 10 orders of magnitude. Note that the material parameters used are close to the experimental measured and targeted values (table \ref{tab2}). We emphasize that the only parameter changed with temperature is the access resistance, where the 2DEG mobility is known to be a strong function of temperature due to electron-phonon interaction in particular \cite{Lisesivdin10}. In the present case, $R$ is adjusted to fit the ohmic regime of the diode above the turn-on voltage and is the single free parameter once the calculation is calibrated at a given temperature.

Figure \ref{Fig7} (\textbf{b}) allows us to go deeper into the transport mechanisms involved under and above the turn-on voltage of the recessed diode by providing the current spectroscopy. As can be seen, the majority of the current at all voltages is emitted in an energy band of a few hundred of meV centered near the Fermi level of the GaN layer. By correlating this observation with band diagrams obtained at different voltages (figure \ref{Fig7}, \textbf{c to e}), different conclusions can be drawn. First, within the voltage range explored, electrons never overcome the AlN potential barrier due to its important height, they tunnel directly through it. Second, below the turn-on voltage (figure \ref{Fig7}, \textbf{c} and \textbf{d}) where the current grows exponentially, hardly any electrons are emitted above the AlGaN conduction band near the AlN/AlGaN interface. At best, they tunnel through its band gap over a certain distance before reaching its conduction band minimum further in the layer. Such a mechanism could be approached to Nordheim-Fowler tunneling \cite{Fowler28}. Overall, within the operating voltage and temperature range, no thermionic processes participate to the transport in the forward conduction regime.  We infer that the use of the thermionic equation for similar devices as it is commonly done, is bound to return incorrect parameters and lead to misinterpretations. 

The derived electronic transport scenario reveals that the recessed TiN/AlGaN/AlN/GaN realizes a tunnel diode  with large output currents. More precisely, the formalism employed suggests that this diode is the stage of coherent electron tunneling events at the vicinity of the anode at room temperature and even higher, a property that would be worth investigating more deeply. From a wider perspective than conventional power conversion, we believe that ultra fast operation could be possible using more compact and fine tuned designs in order to achieve rectification at very high frequencies such as in the infrared spectrum \cite{Davids15}.

\begin{table}[ht]
\centering
\caption{Recessed anode diode main parameters}
\begin{tabular*}{0.45\textwidth}{@{\extracolsep{\fill}}lr}
\hline\hline\\
TiN work function ($W$) & 4.7 eV \\
$n_0$ & 1e15 cm$^{-3}$ \\
$L_{eff}$ & 3 $\mu$m\\
$\omega$ & 26 mm \\
AlGaN Al content & 0.25 (\textit{0.24 measured}) \\
AlGaN thickness & 5.4 nm (\textit{5-7 nm measured})  \\
AlN thickness & 0.58 nm (\textit{1 nm targeted}) \\
$R$ & 0.2 $\Omega$ (300K), 0.37 $\Omega$ (425K)\\\\
\hline\hline
\end{tabular*}
\label{tab2}
\end{table}

Beyond the studied diode design, the sensitivity of electron tunneling to potential barrier heights and thickness raises different questions. First, the removal of the AlN layer would allow higher transparency as far as electrons propagating from the 2DEG to the metal, though it would also imply lower carrier concentration and higher access resistance. Indeed the AlN layer is known to lead to improved sheet resistance in the electron channel due to stronger polarization and enhanced confinement \cite{Smorchkova01}. A comparable argument could be made as far as the Al content in the AlGaN barrier and in both cases a compromise could be possible. Moreover, the reduction of the AlGaN thickness under the TiN metal would also lead to higher transparency. Through further recessing, the impact on the Two-Dimensional Electron Gas arises. For the extreme case of full recess, the 2DEG is removed below the metal gate and a different type of contact is realized: a lateral Metal/GaN contact \cite{Zhu15} for which the transport mechanisms would be worth investigating thoroughly. Finally, for thicker barriers (or alternatively shallower recesses) the coherent tunneling picture might break down due to the loss of phase coherence. Indeed, one cannot rule out incoherent scattering processes over longer tunneling distances and more a general tunneling approach might be required to describe a wider variety of heterostructure designs.

\section{Conclusion and perspectives}
Via an analogy with a MOS contact and current calculations based on the Tsu-Esaki formula, a coherent tunneling transport picture could be provided as far as forward conduction is concerned in a III-N heterojunction diode. This result contrasts the common approach involving thermionic emission which can lead to invalid and misleading physical interpretations. We believe that, in essence, the proposed approach which highlights the importance of field effect and takes into account natively electron emission at multiple energies, can be applied to other architectures and materials. As far as III-N heterojunction  applications are concerned, we may cite RF or power GaN High Electron Mobility Transistors for which the gate electrostatic control and leakage could be tested on a solid basis. We may also mention III-N LED which involve multiple quantum wells and where transport in the forward regime could be explored more thoroughly.

We emphasize that within the coherent tunneling framework used, questions related to trap assisted events or inelastic tunneling may arise \cite{Hill71, Mott69, Ghetti99, Suzuki86, Takagi99}. The good agreement between experimental and calculated data proves that such events do not play an essential role within the device and associated conduction regime explored. Nonetheless pushing the reasoning further to incorporate such effects would provide a larger frame of investigation. Other refinements could be further implemented such as describing the access resistance in better details to capture its temperature variation for example. One may also consider self-heating effects which mainly occur for high current densities and that would open broader voltage ranges to be analyzed. Finally, the energy quantization of the Two-Dimensional Electron Gas, which was treated as a classical incoherent electron reservoir, would be worth evaluating.

\section{Methods}
III-N on Silicon growth was realized via MOCVD using a single chamber closed coupled showerhead equipment manufactured by AIXTRON capable of fully automated handling of 200 mm wafers. Au free process integration was carried out using CEA-LETI's CMOS facilities on the basis of a process flow comprised of approximately 80 technological steps. Electrical characterization was performed by the use of a Tesla 300 mm semi-automatic prober and an Agilent B1505A power device analyzer. All numerical calculations were performed on specifically developed scripts and were double checked using various commercial and open source coding environments.

\section{Acknowledgments}
The authors thank CEA-LETI's clean room department and technical staff for carrying out multiple process integration steps and in-line control measurements. They also thank St\'ephane B\'ecu for fruitful discussions and comments on the manuscript.
This work was performed in the frame of the TOURS 2015 project supported by the french "Programme de l'\'economie num\'erique des Investissements d'Avenir".

\bibliography{Baines_et_al_2017_v4_arxiv}

\end{document}